# Charge-to-Spin Conversion by the Rashba-Edelstein Effect in 2D van der Waals Heterostructures up to Room Temperature

Talieh S. Ghiasi[1]†, Alexey A. Kaverzin[1]†, Patrick J. Blah[1] and Bart J. van Wees[1]

[1]Zernike Institute for Advanced Materials, University of Groningen, Groningen, 9747 AG, The Netherlands

Email: t.s.ghiasi@rug.nl

The proximity of a transition metal dichalcogenide (TMD) to graphene imprints a rich spin texture in graphene and complements its high quality charge/spin transport by inducing spin-orbit coupling (SOC). Rashba and valley-Zeeman SOCs are the origin of charge-to-spin conversion mechanisms such as Rashba-Edelstein effect (REE) and spin Hall effect (SHE). In this work, we experimentally demonstrate for the first time charge-to-spin conversion due to the REE in a monolayer $WS_2$-graphene van der Waals heterostructure. We measure the current-induced spin polarization up to room temperature and control it by a gate electric field. Our observation of the REE and inverse of the effect (IREE) is accompanied by the SHE which we discriminate by symmetry-resolved spin precession under oblique magnetic fields. These measurements also allow for quantification of the efficiencies of charge-to-spin conversion by each of the two effects. These findings are a clear indication of induced Rashba and valley-Zeeman SOC in graphene that lead to generation of spin accumulation and spin current without using ferromagnetic electrodes. These realizations have considerable significance for spintronic applications, providing accessible routes towards all-electrical spin generation and manipulation in two-dimensional materials.

†equal contribution

Spin-orbitronics is a promising field of research that serves the future of spintronic devices, which is based on the manipulation and control of spins and is enabled by spin-orbit coupling (SOC). Graphene is known to be a superior material for long-distance spin transport [1, 2, 3], however, it has intrinsically weak SOC [4]. The control of spin signal that is necessary for spin-based devices becomes possible in graphene by inducing SOC that can be realized via the proximity of materials with large SOC. Recent theoretical [5, 6, 7, 8, 9] and experimental [10, 11, 12, 13, 14, 15, 16, 17] studies have shown that the proximity of transition metal dichalcogenides (TMD) can induce SOC with strength of a few meV in graphene [18]. This leads to a large spin lifetime anisotropy [7, 13, 16], due to the suppression of the in-plane spin lifetime and/or spin absorption [19, 20].

Few orders of magnitude larger SOC in a monolayer TMD [21], compared with graphene, together with its inversion symmetry breaking, provides this semiconductor with theoretically predicted large intrinsic spin Hall angle [22]. Moreover, spin-torque [23] and spin-pumping [24] experiments have shown possibility of charge-to-spin conversion by the Rashba-Edelstein effect in TMDs. However, for the injection/detection and transfer of the spin information, the short spin relaxation time in TMDs presents a major obstacle.

The hybridization of TMD to graphene is an effective way to complement the properties of the these materials. Theory predicts that the band structure of graphene in the proximity of TMD is spin-split by the presence of Rashba and valley-Zeeman spin-orbit fields [5, 6]. These spin-orbit fields are the origin of charge-to-spin conversion mechanisms such as the Rashba-Edelstein effect (REE) and the spin Hall effect (SHE) that generate spin accumulation and spin-polarized currents, respectively [9, 25, 26, 27, 28, 29, 30]. More importantly, the strength of these spin-orbit fields, and so the efficiency of the charge-to-spin conversion mechanisms, is dependent on the position of the Fermi-energy within the band structure of the TMD-graphene heterostructure. For the first time, we show in this work that this is indeed the case for a $WS_2$-graphene heterostructure where the Rashba-Edelstein effect, in particular, creates a spin accumulation within the graphene channel which is detectable up to room temperature, and it is largely tunable by a gate transverse electric field.

The Rashba SOC in graphene originates from breaking the out-of-plane symmetry due to the proximity of the TMD [26]. The resulting out-of-plane effective electric field ($\mathbf{E} = E\hat{\mathbf{z}}$) generates an in-plane Rashba spin-orbit field ($\sim \mathbf{E} \times \mathbf{p}$) that is perpendicular to the momentum ($\mathbf{p}$) of the electrons in the Dirac cone and ultimately creates a tangential winding spin texture of the electron states in momentum space. Due to the Rashba-Edelstein effect, a charge current (density) $\mathbf{J}$ generates a non-zero spin density ($\propto \hat{\mathbf{z}} \times \mathbf{J}$), polarized perpendicular to the current direction [25, 26].

The winding Rashba spin-orbit field in the graphene changes sign between the spin-split Dirac cones of the conduction (or valence) band (Figure 1a). Therefore, the current-driven spin densities of the spin-split bands have opposite sign, which reduces the total spin density at the Fermi energy. However, the energy gap between the spin-split Dirac cones is enhanced by the presence of the valley-Zeeman field (calculated about $2.2\,\mathrm{meV}$ for $WS_2$-graphene [6]). This results in considerably different magnitudes of the (current-driven) spin densities associated with each of the cones for low-energy states. This avoids compensation of the spin accumulation from the bands with opposite spin-winding and therefore it helps to optimize the efficiency of the charge to spin conversion [9].

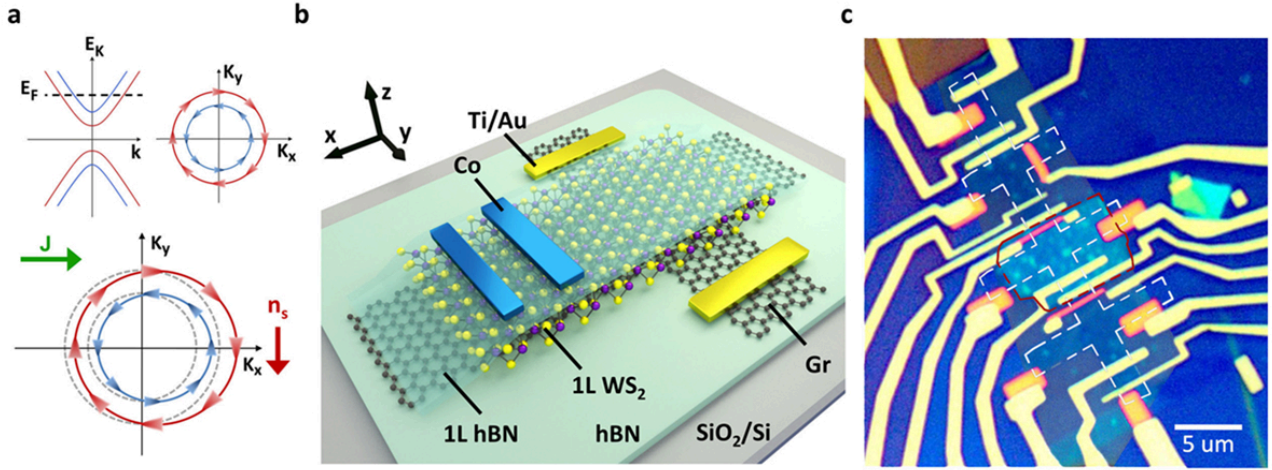

**Figure 1.** (a) TMD-graphene band structure, consisting of spin-split Dirac cones with opposite spin helicity. The charge current (J), shifts the Fermi level contours from equilibrium (gray-dashed lines) and induces a non-equilibrium spin density ($n_s$) by the Rashba-Edelstein effect. (b) Sketch of the van der Waals heterostructure of 1L WS$_2$-graphene, encapsulated with the top monolayer hexagonal Boron Nitride (hBN) and bottom bulk hBN (with the thickness of 14 nm). The device is made with Ti/Au and Co electrodes on a SiO$_2$/doped Si substrate. The sketch illustrates the central region of the sample including the electrodes that are used for our measurements. (c) Optical microscope image of the fabricated device. The red and white dashed lines show the edges of the WS$_2$ and etched graphene flake, respectively.

We observe clear evidence of the charge-to-spin conversion in TMD-graphene heterostructures due to the REE, which is accompanied by the SHE. Different directions of the spins generated from these two effects make their contributions distinguishable by their distinct symmetries as a function of magnitude and direction of the magnetic field in our (oblique) Hanle precession measurements. SHE is recently observed in multilayer MoS$_2$/multi-layer graphene [30], where the SHE signal in graphene is superimposed by an additional spin-to-charge conversion mechanism which is mainly associated to SHE in the bulk MoS$_2$. However, the measurements in this work are performed on a vdW heterostructure of a single layer of WS$_2$ and graphene. The two-dimensionality of monolayer (1L) TMD compared to bulk TMD [31] eliminates the vertical charge transport inside the 1L TMD. Therefore a possible contribution from the SHE in bulk TMD is largely suppressed in our system. Stronger induced SOC in graphene by 1L TMD, as compared with bulk [17], in addition to the theoretical prediction of the largest SHE signal, specifically, in a 1L WS$_2$-graphene heterostructure [29], makes the vdW stack of our sample an optimal choice.

In Figure 1b, we show the device geometry consisting of 1L WS$_2$/1L graphene that is encapsulated between 1L hexagonal Boron-Nitride (hBN) and bulk hBN. The device is fabricated on a 300 nm SiO$_2$/doped Si substrate with Ti/Au and Co electrodes, made by shadow mask evaporation and e-beam lithography, respectively (see Methods). The 1L hBN acts as a tunnel barrier for the spin injection/detection by the Co electrodes. Represented by the white dashed line in the optical image (Figure 1c), the graphene channel is etched into a Hall-bar which allows for the non-local detection of the induced spin density, generated by both effects. Note that for fabrication of the vdW stack, we do not have control over the crystallographic alignment of the TMD, graphene and hBN flakes which can affect the strength of the spin-orbit fields as compared with the calculations that assume (super)lattice matching [32].

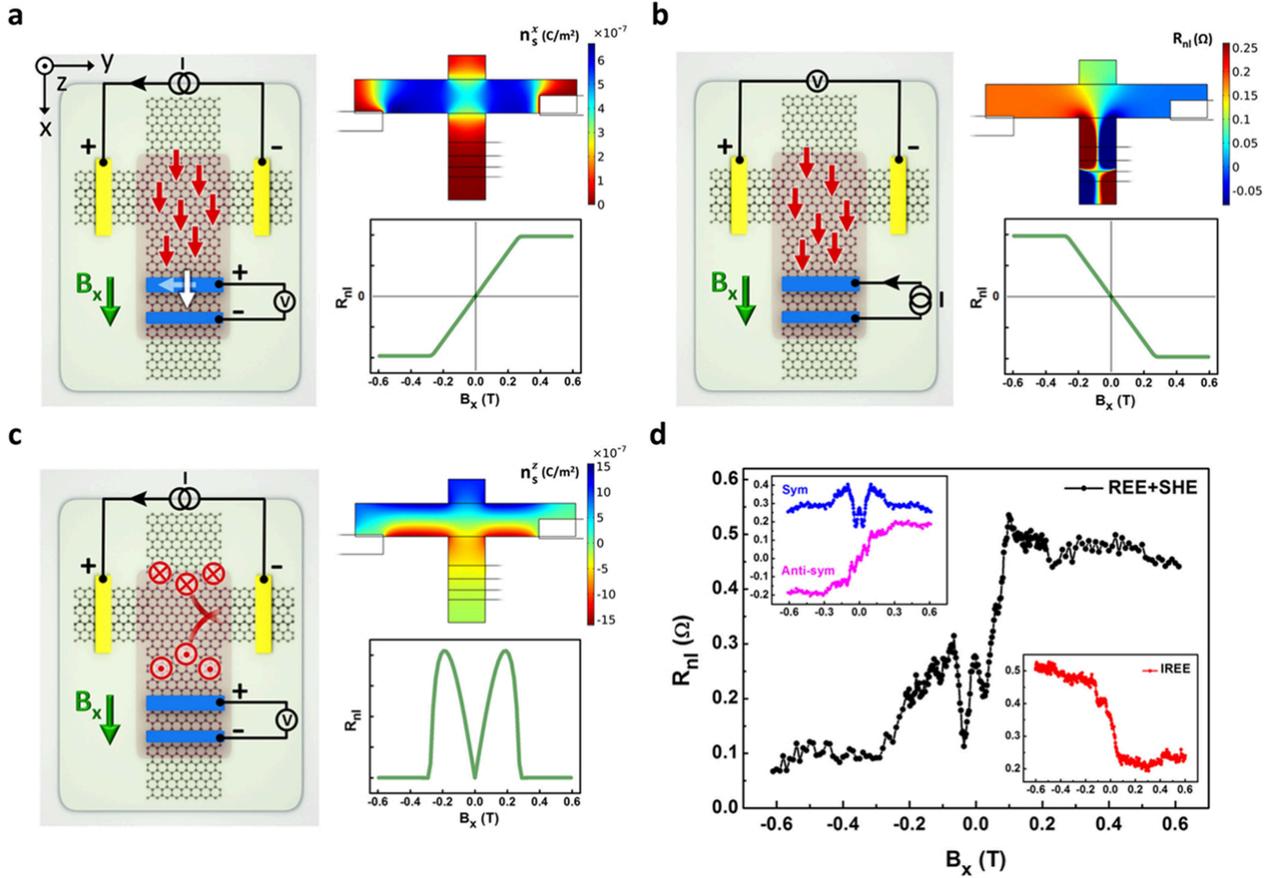

**Figure 2. Rashba-Edelstein effect (REE), inverse REE (IREE) and spin Hall effect(SHE)**. (a) Device sketch and measurement geometry for REE. A current source (I) is applied to the Ti/Au electrodes and the voltage (V) is measured across the Co electrodes. Red arrows represent the accumulation of in-plane spins polarized along the x-axis. The colormap is the magnitude of the spin density polarized along the x-axis ($n_s^x$) at zero magnetic field ($B_x = 0$ T), over the sample. The plot shows the modulation of the non-local resistance ($R_{nl}$) vs $B_x$, calculated considering the Stoner-Wohlfarth model for the behavior of the Co contacts. The white arrows in the sketch represent the Co magnetization direction. (b) Inverse REE (IREE) measurement geometry, colormap for the voltage distribution at $B_x = 0$ T and the corresponding expected modulation of the $R_{nl}$ vs $B_x$. (c) Measurement geometry for SHE and generation of out-of-plane polarized spin current, resulting in accumulation of out-of-plane spins (polarized along the z-axis). The colormap shows the density of the out-of-plane spins ($n_s^z$) at $B_x = 0$ T. The plot shows the modulation of $R_{nl}$, expected from Hanle precession of the out-of-plane spins, resulting in symmetric behaviour vs $B_x$ (considering the y-component of the Co contact magnetization is orienting according to the y-component of the magnetic field direction). (d) $R_{nl}$ versus $B_x$, experimentally measured in the geometry of the REE and SHE (with $I = 5\,\mu$A). The measurement is performed at 4.2 K with charge carrier density of $+1.5 \times 10^{12}\,\mathrm{cm}^{-2}$. The inset at the top-left shows the symmetric and anti-symmetric components of the signal, separately. The inset at the bottom-right is the $R_{nl}$ measured in IREE geometry (with $I = 2\,\mu$A).

Our main focus in this work is on the TMD-covered graphene region of this device. As shown in the device sketch of Figure 2a, using the Ti/Au contacts on graphene we apply charge current ($I$) and with ferromagnetic Co electrodes we measure the non-local voltage ($V_{nl}$) as a function of an applied magnetic field (B). With an applied charge current along the y-axis and in the presence of REE one should expect generation of non-zero spin density polarized along the x-axis, $n_s^x$. We formulate our theoretical model of coupled charge-spin transport in the presence of REE. By numerically solving Bloch diffusion equations (COMSOL, see SI for details), we obtain a distribution of $n_s^x$ over the full sample shown as a colormap in Fig 2a. Using these solutions at any applied field $B$ one can estimate the signal between the spin-sensitive Co contacts, shown in the bottom right corner of Figure 2a.

At $B = 0\,\text{T}$ the magnetization of the Co contacts is along their easy axis (y-direction) implying that the non-local resistance ($R_{nl}$) should be zero. Applying a magnetic field in x-direction ($B_x$) changes the direction of the contact magnetization, in accordance with the Stoner-Wohlfarth model [33]. The component of the contact magnetization along the x-axis increases linearly with $B_x$ while the REE-induced spin density stays unaffected. This results in a linear increase of the non-local resistance until the contact magnetization direction is completely saturated along the x-axis (at $B_x \sim 0.3\,\text{T}$, for our Co electrode geometry). A negative magnetic field causes the alignment of the contact magnetization towards the opposite direction. This results in a negative signal since the polarization of the REE spin density stays unchanged. Therefore, anti-symmetric behaviour of the non-local signal versus $B$ is a direct signature of REE and can be used to extract the REE related signal from the experimental results.

By the inverse of the REE (IREE), generation of charge current becomes possible as a result of the non-zero spin density in graphene [34]. In this geometry (shown in Figure 2b), the detection of the non-local voltage drop is across the Ti/Au contacts, while Co electrodes are used to apply the current required for injection of in-plane spins. IREE is the Onsager reciprocal of REE implying that the detected non-local signal should be the same but with a reversed sign of $B$ ($R_{ij,kl}(B) = R_{kl,ij}(-B)$, with ij and kl are the indices of current and voltage terminals).

In Figure 2c, we illustrate the mechanism for creating spin current and the resulting accumulation of spins by SHE. In this case a spin current with out-of-plane polarization is generated perpendicular to the direction of the charge current [35]. In the colormap of figure 2c, we show the out-of-plane spin density ($n_s^z$) produced by the SHE all over the sample. The out-of-plane spins cannot generate a non-local voltage across the in-plane magnetized Co electrodes, unless they precess around the applied magnetic field. Therefore, the detected signal develops from zero at $B = 0\,\text{T}$ to a finite value as the spins precess to the in-plane direction along the y-axis. Furthermore, the $R_{nl}$ drops back to zero above the saturation field of the Co contact as the Co magnetization and spin alignment are again perpendicular to each other. The sign of the SHE signal depends on the orientation of the Co magnetization. In our calculations we assume that the y-component of the contact magnetization is orienting according to the y-component of the magnetic field direction (see SI for details). This results in a symmetric behaviour of SHE component versus $B$ which is thus easily distinguishable from the anti-symmetric REE component.

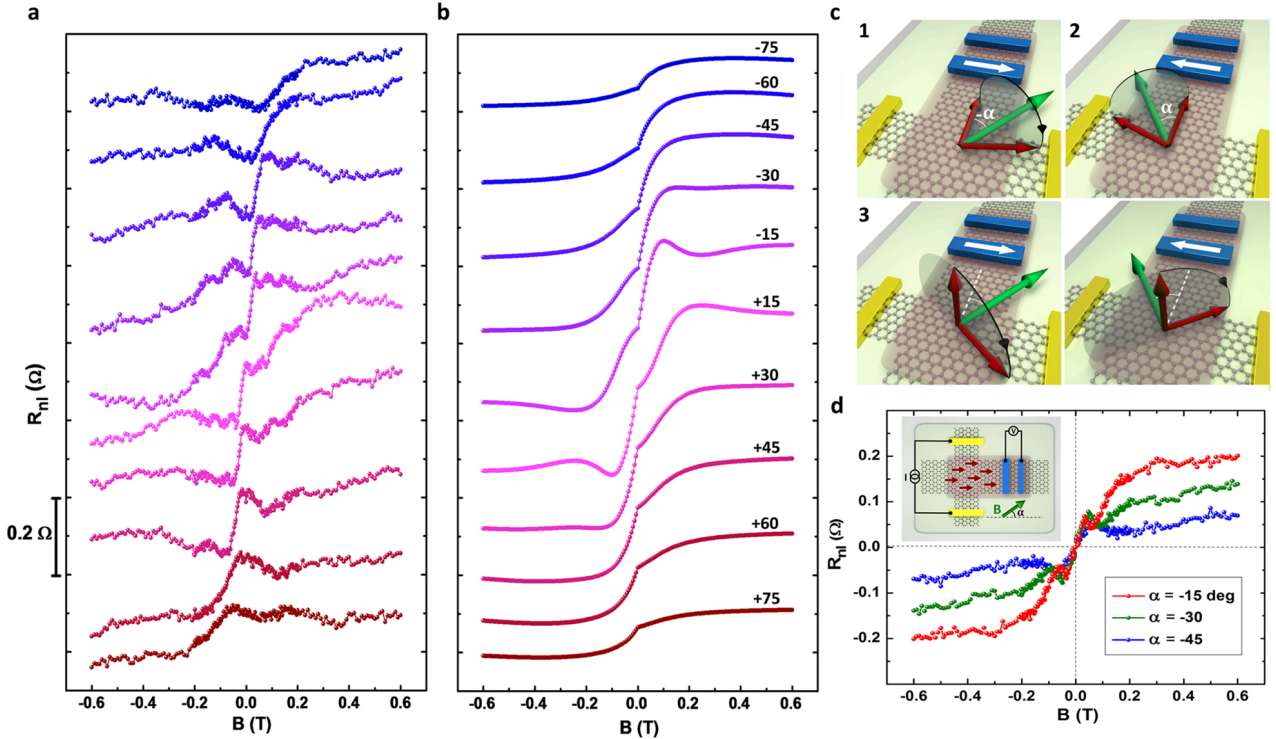

**Figure 3. Precession of REE and SHE spins under $\alpha$-angled magnetic field.** (a) $R_{\mathrm{nl}}$ measured versus $B$ applied under angle of $\alpha \approx$ -75 to +75 degrees. The unknown component is subtracted from this data set. (b) closest fit to the data, obtained for $\tau_{\parallel} \approx 3.5$ ps and $\tau_{\perp} \approx 90$ ps, considering the spin hall angle ($\Theta_{\mathrm{SH}}$) of 0.13 and REE conversion efficiency ($\alpha_{\mathrm{RE}}$) of 2.8 (defined as the ratio of spin density over charge current density $n_{\mathrm{s}}^x/(2v_{\mathrm{F}}J_y)$, with $v_{\mathrm{F}}$ as Fermi velocity).(c) Device sketch with symmetries of Hanle signal vs angle while considering the precession of the in-plane spins (panel 1 and 2) and the out-of-plane spins (panel 3 and 4). (d) Measurement geometry for REE. Anti-symmetric component of $R_{\mathrm{nl}}$ versus B, applied under angles of $\alpha \sim 15$, 30 and 45 degrees.

We show our experimental result in Figure 2d, obtained by applying a current source of $5\,\mu$A and measuring the $R_{\mathrm{nl}}$ by the Co electrode located at $2\,\mu$m from the center of the graphene cross (graphene width is also $2\,\mu$m). These measurements are performed at 4.2 K with charge carrier density of $+1.5\times10^{12}\,\mathrm{cm}^{-2}$. The observed result contains signals from both REE and SHE effects. The top-left inset shows the anti-symmetric and symmetric components that are extracted from the measured data in order to discriminate the spin signal dominated by the REE and SHE, respectively. The magnitude of the measured

REE spin signal is $\Delta R_{\mathrm{nl}} \approx 200\,\mathrm{m}\Omega$, defined as half of the difference between the $R_{\mathrm{nl}}$ values measured at the two saturation levels.

The bottom-right inset shows the IREE spin signal measured with the inverse geometry that shows similar behaviour, however with reversed sign versus $B$, confirming the spin-to-charge conversion and preservation of the reciprocity in the linear regime. The very small background resistance in these measurements affirms that, in our non-local geometry the current path is well-separated from the voltage probes so that any spurious effect can be dismissed.

The magnitude and modulation of the

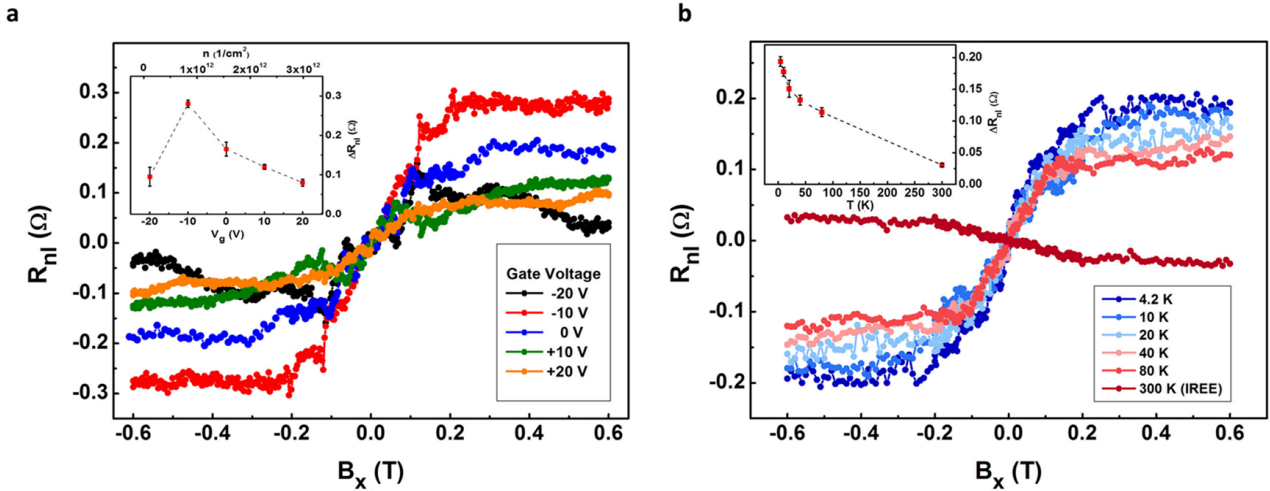

**Figure 4. Gate and temperature dependence of REE spin signal.** (a) Hanle precession measured with respect to $B_x$ (anti-symmetrized $R_{nl}$ vs $B$) at gate voltages of -20 V to +20 V. The inset is the magnitude of REE spin signal versus $V_g$. (b) anti-symmetric component of $R_{nl}$ vs $B_x$, measured at different temperatures in the REE geometry. The signal shown at room temperature is measured in IREE geometry. The inset is the temperature dependence of the REE spin signal.

measured spin signal is strongly dependent on the direction of the applied magnetic field. In Figure 3a we evaluate how the non-local resistance changes as we apply the in-plane magnetic field under certain angles with respect to the x-axis ($\alpha \approx$ -75 to +75 deg). All the measurements are performed by aligning the contact magnetization at high fields, meaning that the y-component of the Co magnetization is always collinear with the y-component of the magnetic field. In panel b, we show the corresponding modeled dependences that closely reproduces our experimental results.

The behaviour of the $R_{nl}$ is understood by considering the precession of the in-plane and out-of-plane spins around the $\alpha$-angled magnetic field and the corresponding symmetries versus $\alpha$. As shown in Figure 3c, REE induced spins result in the same positive projection on the Co magnetization direction for both $+\alpha$ and $-\alpha$, meaning that the REE spin signal is symmetric versus the angle $\alpha$. On the other hand, the precessed out-of-plane SHE spins generate signal projections on the contact magnetization with opposite sign for $+\alpha$ and $-\alpha$ implying anti-symmetric contribution of SHE spins versus the angle. This means that the REE contribution to the signal does not change whereas the SHE contribution changes from peaks(dips) to dips(peaks) when the angle is changed from $+\alpha$ to $-\alpha$.

Specifically, in Figure 3d we show the REE spin signal (anti-symmetric vs $B$ and symmetric vs $\alpha$) measured under angled B. As expected for the Co magnetization behaviour, we observe the shift of the saturation fields under different angles, together with the change in the magnitude of the spin signal. In the following table we summarize the symmetries for the in-plane Hanle precession measurements as:

| vs | REE | SHE |
|---|---|---|
| $B$ | anti-sym | sym |
| $\alpha$ | sym | anti-sym |

Note that in the measured non-local signal, there is an additional component that does not comply with the symmetries of the REE and SHE. This component is subtracted from the experimental data, resulting in Figure 3a (for details and discussions see SI).

The closest fit to the data (Figure 3b) gives an estimate for the in-plane spin lifetime of $\tau_\parallel \approx 3.5$ ps with a spin lifetime anisotropy of $\tau_\perp/\tau_\parallel \approx 26$, considering the spin hall angle ($\Theta_{SH}$) of 0.13 and REE conversion efficiency ($\alpha_{RE}$) of 2.8 (defined as the ratio of spin density over charge current density $n_s^x/(2v_F J_y)$, with $v_F$ as the Fermi velocity, See Ref. [9] and SI for details). We formulate the spin Hall angle and the REE efficiency as $\Theta_{SH} = SH\rho$ and $\alpha_{RE} = 2RE\,v_F\rho\tau_\parallel$, where $\rho$ is the resistivity of the TMD-covered graphene channel and the $SH$ and $RE$ are the strengths of the two effects. The ratio between the strengths is defined as $SH/RE = 2\Theta_{SH}v_F\tau/\alpha_{RE} \simeq 0.33\,\mu m$. Since in our analysis the Co contact polarizations cannot be extracted independently from the REE and SHE strengths, the ratio between the strengths is more accurate compared to their individual values. For comparison, the reported value for $\Theta_{SH}$ in bulk $MoS_2$-graphene is about 0.05 (with undefined charge carrier density) [30] .

One of the most important requirements for the spin-based devices is the possibility for tuning the spin signal by a gate electric field. Here we demonstrate the modulation of the REE efficiency with a gate. The REE is theoretically predicted to be gate-tunable [9] due to its strong dependence on the spin-split band structure of TMD-graphene heterostructure. We evaluate this by comparing it to our measurements in REE geometry at different back-gate voltages ($V_g$), shown in Figure 4a. Electrical characterization of the graphene channel shows n-type doping with the charge neutrality point at $V_g = -22$ V (see SI). Measurements performed close to the Dirac point are difficult to interpret due to presence of inhomogeneity originating from disorder. In addition, in this regime the contact resistance becomes comparable with channel resistance, which can suppress spin transport considerably. Therefore, we exclude measurements performed at the Dirac point from our consideration. However, we observe that the increase of the gate voltage from -10 V to +20 V (change in Fermi energy from 100 meV to 200 meV) results in a considerable decrease of the spin signal, ~70 percent.

This behaviour can be associated with the fact that $V_g$ shifts the Fermi energy from the charge neutrality point into the conduction band, at which both of the spin-split Dirac cones (with opposite spin-winding directions) are available. The opposite winding of the spin texture of the two bands reduces the efficiency of the REE to a large extent, leading to lower in-plane spin density. We observe that the measured REE spin signal decays as a function of gate-voltage which is in agreement with the theoretically predicted decay in the REE efficiency versus position of the Fermi energy [9].

The preservation of the charge-to-spin conversion mechanism at room-temperature is a prerequisite for potential applications. We evaluate the temperature dependence of the REE (Figure 4b) and observe that the spin signal generated by the REE and the IREE is preserved up to room temperature, however it decays by about 80 percent from 4 k up to RT. This behaviour indicates the robustness of the REE charge-to-spin conversion mechanism, which is in agreement with theoretical predictions [9]. We observe that the features associated with SHE in our system, together with the unknown component in the Hanle precession measurements, vanish at temperatures above 20 K. This indicates that the SHE has a stronger dependence on temperature compared to the REE.

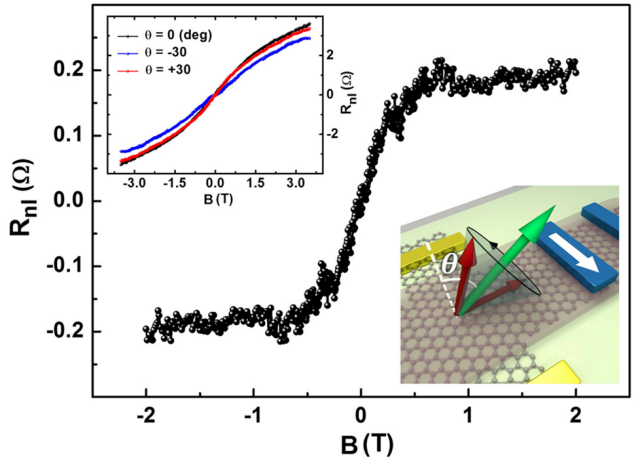

**Figure 5. Observation of REE by out-of-plane Hanle precession measurements.** The device sketch shows the precession of the in-plane REE spins, about the applied magnetic field, angled by $\theta$ with respect to the normal to the plane. The curve is the anti-symmetric component of the subtraction of the signal measured under angles of $\theta = \pm 30$ deg (attributed to REE spin signal). The inset is the anti-symmetric component of the nonlocal resistance as a function of B, measured with respect to the $\theta$-angled magnetic field.

We further characterize the spin transport by applying the magnetic field in the x-z plane, under an angle ($\theta$) with respect to the normal to the sample plane (shown in the sketch of Figure 5). Firstly, the $\theta$-angled $B$ brings the contact magnetization direction out-of-plane and secondly, it precesses the in-plane spins into the out-of-plane direction, which are then detected by the contacts with a tilted magnetization. The symmetry table for both SHE and REE components for the out of plane field measurements is:

| vs | REE | SHE |
|---|---|---|
| $B$ | anti-sym | anti-sym |
| $\theta$ | anti-sym | sym |

SHE is symmetric vs angle and is anti-symmetric vs $B$, thus having the same symmetries as the ordinary Hall effect. The non-local sample geometry minimizes the local charge current contribution to the detection voltage, to a large extent. However, the detection electrodes in our sample are close to the current path which results in a (small) Hall effect contribution of a similar order as SHE contribution. This implies that we are not able to extract the SHE component. Nevertheless, by anti-symmetrizing the measured signal with respect to the angle we eliminate both SHE and regular Hall effect, thus, leaving only REE contribution.

In Figure 5 we show the resulting dependence of the described procedure which gives $\sim 200$ m$\Omega$ of REE spin signal, measured at RT with carrier density of $-3.6 \times 10^{11}$ cm$^{-2}$. Note that this is considerably larger than the REE associated spin signal obtained via the in-plane geometry, which is consistent with an increase in REE efficiency with lower carrier density.

In this work, we also observe modulations of the second harmonic signal ($V_{nl}/I^2$). The results show considerable dependence on the applied magnetic field and gate electric field which is a signature of thermally driven spin polarization in the TMD-graphene heterostructure. This observation (results are shown in SI) could be an indication of spin-Nernst or spin-Seebeck effects [36] in this system, however, it requires further studies.

Our experimental observations are unambiguous evidence for the presence of both Rashba-Edelstein and spin-Hall charge-to-spin conversion mechanisms in a monolayer TMD-graphene heterostructure. This is the direct proof of the effective imprint of the Rashba and valley-Zeeman spin-orbit fields in graphene, while its charge transport properties are pre-served. In this work, we comprehensively addressed the charge-induced non-equilibrium spin density, generated by the REE and we employed strategies in order to discriminate this effect from SHE by symmetries of the Hanle precession measurements as a function of oblique magnetic fields. The ability to address the individual effects in one heterostructure, allows for a valid comparison of their strengths. Moreover, the observed strong dependence of the REE spin signal on the position of Fermi energy shows efficient tunability of spin generation by a transverse electric field. This observation in addition to the fact that the spin signal remains considerable up to room temperature confirms that the monolayer TMD-graphene heterostructure is a promising choice for the future of two-dimensional spin transistors without the need for bulk ferromagnetic electrodes.

## Methods

*Device Fabrication.* The monolayers of WS$_2$ and graphene and h-BN (1L and bulk) are mechanically cleaved from their bulk crystals (provided by HQ graphene) on SiO$_2$/Si substrates, using adhesive tapes [37]. The monolayer flakes are identified by their optical contrast with respect to the substrate [38]. Thicknesses of the flakes are verified by atomic force microscopy (AFM). Using a dry pickup technique [39], we transfer the graphene on the bulk hBN flake. By a pre-patterned PMMA mask, we etch the graphene-bulk hBN by oxygen-plasma into a H-bar geometry. We finalize the fabrication of the vdW stack by the transfer of the 1L hBN-WS$_2$ on top of the etched graphene-bulk hBN. We proceed with fabrication of electrodes on the vdW stack by shadow-mask evaporation and e-beam lithography technique (using PMMA as the e-beam resist). Due to the complication of the fabrication process, there is a high chance for breaking the graphene channel. In the sample studied in this work, the presence of few cracks in our graphene channel has made the TMD-covered graphene region, electrically disconnected from the rest of the sample. Therefore, our analysis is focused only on the TMD-covered graphene region, shown in the device sketch of Figure 1b.

*Electrical Measurements.* The charge and spin transport measurements are performed by using standard low-frequency ($< 20$ Hz) lock-in technique with AC current source of 100 nA to 5 $\mu$A. A Keithley source-meter is used as the DC-voltage source for the gate. Rotatable sample stages (separate for the in-plane and out-of-plane measurements) are used for applying the magnetic field by a (superconducting) magnet in all the possible directions.

## Acknowledgements

We kindly acknowledge Josep Ingla-Aynés for insightful discussions and we thank T. J. Schouten, H. Adema, Hans de Vries and J. G. Holstein for technical support. This research has received funding from the Dutch Foundation for Fundamental Research on Matter (FOM) as a part of the Netherlands Organisation for Scientific Research (NWO), FLAG-ERA (15FLAG01-2), the European Unions Horizon 2020 research and innovation programme under grant agreements No 696656 and 785219 (Graphene Flagship Core 1 and Core 2), NanoNed, the Zernike Institute for Advanced Materials, and the Spinoza Prize awarded in 2016 to B. J. van Wees by NWO.

# Supporting Information:
# Charge-to-Spin Conversion by the Rashba-Edelstein Effect in 2D van der Waals Heterostructures up to Room Temperature

Talieh S. Ghiasi[1]‡, Alexey A. Kaverzin[1]‡, Patrick J. Blah[1] and Bart J. van Wees[1]

[1]Zernike Institute for Advanced Materials, University of Groningen, Groningen, 9747 AG, The Netherlands

Email: t.s.ghiasi@rug.nl

‡equal contribution

## 1. Sample fabrication and AFM characterization

Fabrication of the fully hBN-encapsulated van der Waals (vdW) heterostructure of TMD-graphene (Figure 1a) starts with separate exfoliation of the TMD, graphene and hBN on $SiO_2$/Si substrates. Optical contrast of the flakes with respect to the substrates, together with the height profiles, measured by AFM, confirms that the selected flakes are monolayers. For fabrication of the vdW stacks we use a dry pick-up technique [1], using poly(carbonate) and PDMS stamps. First, we transfer the monolayer (1L) graphene on the bulk hBN flake. In order to release the graphene flake, we melt the PC at 190 °C and we remove the PC in chloroform for 5 min. Further, the sample is annealed at 350 °C in $Ar/H_2$ atmosphere for 4 hrs. In the next step, we etch the graphene-bulk hBN stack into Hall-bar geometry by oxygen plasma, using a lithographically prepared PMMA mask (details of this technique will be explained in a separate publication).

Using another PC-PDMS stamp, we pick up the 1L hBN from the $SiO_2$/Si substrate that it is exfoliated on. The 1L hBN on PC-PDMS stamp is brought in contact with 1L TMD for pick up by using van der Waals forces. The 1L hBN/1L TMD stack on the PC-PDMS stamp is placed on top of the etched graphene-bulk hBN stack on $SiO_2$/Si substrate by micro-manipulators. The process of melting the PC and its removal in chloroform and further annealing in $Ar/H_2$ atmosphere is repeated, accordingly. The procedure for the full hBN-encapsulation of TMD-graphene heterostructure is followed by the fabrication of electrodes. The Ti/Au electrodes are made by using a pre-patterned (by EBL) PMMA mask. By E-beam evaporation of Ti/Au, we make the gold pads. The Co electrodes are fabricated by conventional EBL technique, using PMMA as the e-beam resist, followed by the deposition of Co at UHV atmosphere and lift-off in 40 °C acetone.

Figure 1b shows an Atomic Force Microscope (AFM) image (tapping mode) of the fully hBN-encapsulated 1L TMD-graphene on $SiO_2$/Si substrate. The height profile of each of the layers is shown in the inset, which confirms consistency with thicknesses associated with monolayers of TMD, graphene and hBN. In the AFM image, we observe formation of bubbles at the interface between the 2D crystals. These bubbles that mainly originate from adsorbate molecules, indicate proper adhesion between layers of the vdW heterostructure, meaning that the area in between the bubbles is expected to be clean and adsorbate-free.

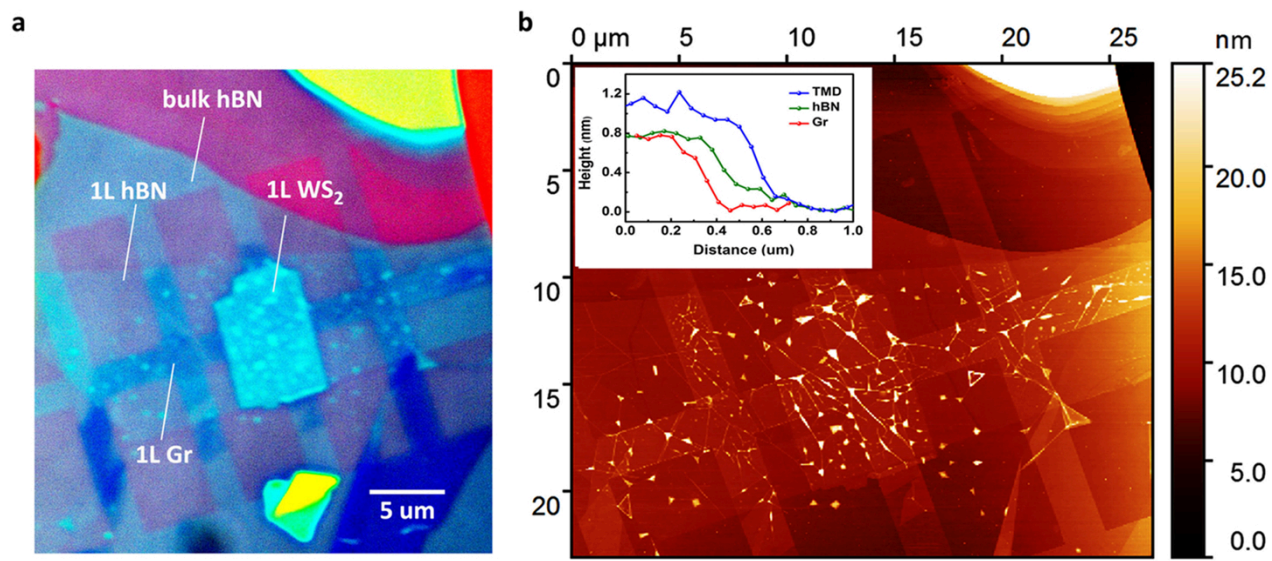

**Figure 1. Optical and atomic force microscope (AFM) images** (a) Optical microscope image of the hBN-encapsulated WS$_2$-graphene heterostructure. (b) AFM image and height profile (inset) of the vdW stack (tapping mode). The thickness of the bottom hBN is about 14 nm.

## 2. Charge transport measurements

We perform four-terminal electrical measurements in order to characterize the charge transport properties of the TMD-graphene heterostructure. In Figure 2, we show the gate-dependence of the conductance measured through the TMD-covered graphene region. In order to measure the spin transport with magnetic field applied in and out-of-plane, discussed in this work, we had to bring the sample to ambient condition and change the sample rotator. In Figure 2, panel (a) shows the transport characteristics of the channel, for the measurements with an in-plane rotator (B applied in x-y plane) and panel (b) shows that of the out-of-plane rotator (B applied in x-z plane). We observe that the doping of the channel changes from n-doping to p-doping during the sample transfer. From the charge transport, considering the geometry of the channel, we extract the charge diffusion coefficient, used for our further analysis for spin transport (assuming $D_c = D_s$). The charge transport characteristics of the sample shows that the TMD-covered graphene region that is the focus of this work is electrically disconnected from the rest of the sample. Therefore, we restrict our measurements to the TMD-covered region as shown in Figure 1(b) of the manuscript.

The resistance of the contacts is estimated from a three-probe measurements with subtraction of channel resistance contribution (assuming homogenous channel). The resistances of the Ti/Au contacts are $\sim 3$ and 5 k$\Omega$ and those of the Co contacts are $\sim 4$ and 14 k$\Omega$. The larger resistance of the contacts compared with that of the graphene channel suppresses the back-flow induced spin relaxation.

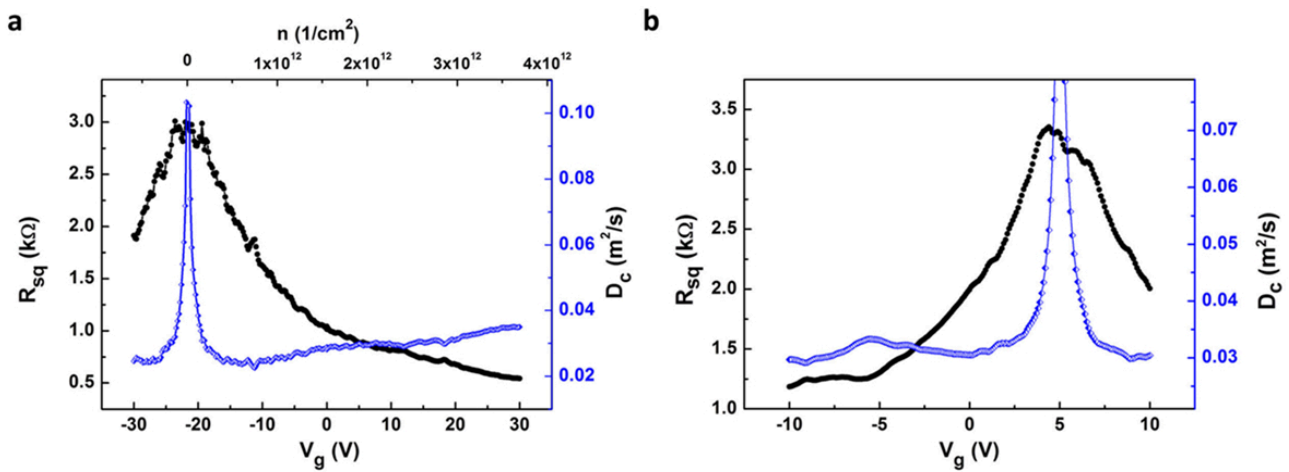

**Figure 2.** TMD-graphene charge transport characteristics. Square resistance ($R_{sq}$) and diffusion coefficient ($D_c$) of the WS$_2$-covered graphene channel versus gate-voltage ($V_g$) and density of charge carriers measured (a) in the in-plane rotator and (b) in the out-of-plane rotator.

### 3. Symmetrization and antisymmetrizations of Hanle precession measurements vs direction and magnitude of the in-plane magnetic field B

As discussed in the main manuscript, SHE and REE are expected to produce signals that obey specific symmetries with respect to the strength of magnetic field B and in-plane angle $\alpha$ under assumption that the ferromagnetic contact has opposite magnetization alignment for positive and negative B. The details on the magnetization alignment of contacts with respect to $B$ are discussed in section 4. The symmetry table for the SHE and REE components is given in the main text. According to the table (anti-)symmetrizing the data with respect to $B$ should already disentangle the REE and SHE contribution in the measured signal. However, we see from our analysis that the extracted symmetric vs $B$ component of the data is neither symmetric nor anti-symmetric with respect to $\alpha$ whereas the SHE component should be purely anti-symmetric in $\alpha$. This means that in our results we have an additional component that cannot be explained either by REE or SHE. Therefore, we perform further (anti-)symmetrization of our data with respect to $\alpha$ in order to extract purely REE and SHE associated components. REE is extracted by anti-symmetrizing the original data vs $B$ and symmetrizing it vs $\alpha$. SHE is extracted by symmetrizing vs $B$ and anti-symmetrizing vs $\alpha$. In the main manuscript we show in Figure 2 only the sum of the SHE and REE related contributions. The original full set of data is shown here in Figure 3 plotted for different angles $\alpha$, after performing alignment of the Co contact magnetization at $\pm 3\,\mathrm{T}$ and sweeping the magnetic field sweep from $\pm 0.6\,\mathrm{T}$ to $0\,\mathrm{T}$, respectively.

In panels of Figure 4(a)-(i) we illustrate the decomposition procedure implemented here for the curve measured at $\alpha = -15$ deg. In panel (a) the measured data is shown. By anti-symmetrising and symmetrising it with respect to the field we obtain components shown in panels (d) and (g), respectively. Further, the obtained curves can be symmetrized and anti-symmetrized with respect to angle by using a corresponding curve measured at $\alpha = +15$ deg (not shown in this figure). The resulting decomposition (panels (e), (f), (h), (i)) gives a contribution coming from SHE (symmetric in B, anti-symmetric in $\alpha$) and REE (anti-symmetric in B, symmetric in $\alpha$) plus another two contributions (symmetric in B, symmetric in $\alpha$(panel i) and anti-symmetric in B, anti-symmetric in $\alpha$(panel h)) the origin of which is not understood. The sum of SHE and REE components is plotted in panel (d) and gives the data that was shown in the main manuscript (Figure 3a) and was used to compare with the model outcome. Note that the panel (g) shows the total unknown contribution that is still smaller in the magnitude compared to the REE component which is the main focus of this work.

One of the possible explanations of the unknown component could be the generation of $y$-axis aligned spins due to misalignment of charge current in our sample. However, $n_s^y$ produces the contribution to the measured signal that should also obey symmetries according to the symmetry table as we summarise here:

| vs | $n_s^x$ (REE) | $n_s^z$ (SHE) | $n_s^y$ |
|---|---|---|---|
| B | anti-sym | sym | anti-sym |
| $\alpha$ | sym | anti-sym | anti-sym |

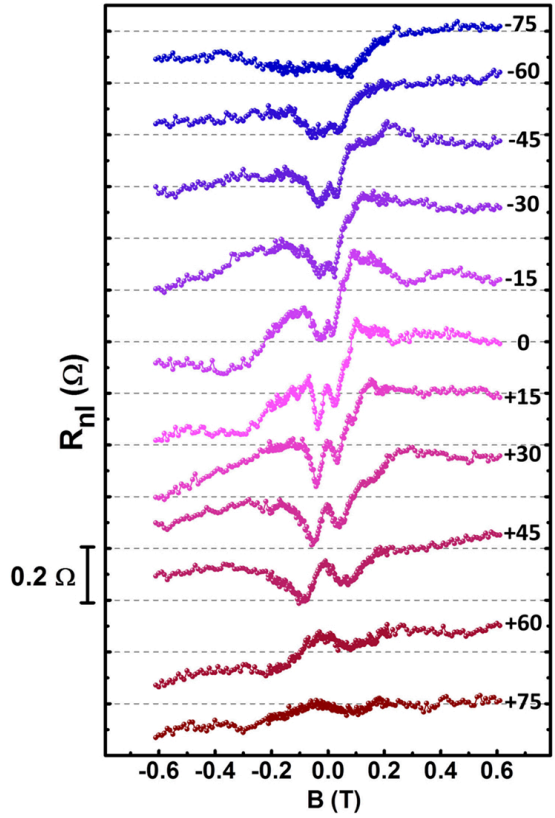

**Figure 3.** Nonlocal signal measured as a function of an in-plane magnetic field at different angles $\alpha$ from $-75$ to $+75$ degrees. Prior to the measurements in the positive/negative field range the Co contacts magnetizations are aligned by applying $+3\,\mathrm{T}/-3\,\mathrm{T}$.

This means that presence of $n_s^y$ should result in anti-symmetric vs $B$ behaviour whereas we see the largest unknown contribution in a symmetric vs $B$ component of the measured signal. Another possible explanation is a non-homogeneous distribution of the magnetization over the ferromagnetic contact which can result in a signal that does not obey symmetry with respect to the angle $\alpha$. Therefore, we assume the non-ideal behavior of Co contacts as the most probable explanation of the unknown contribution to our measurements.

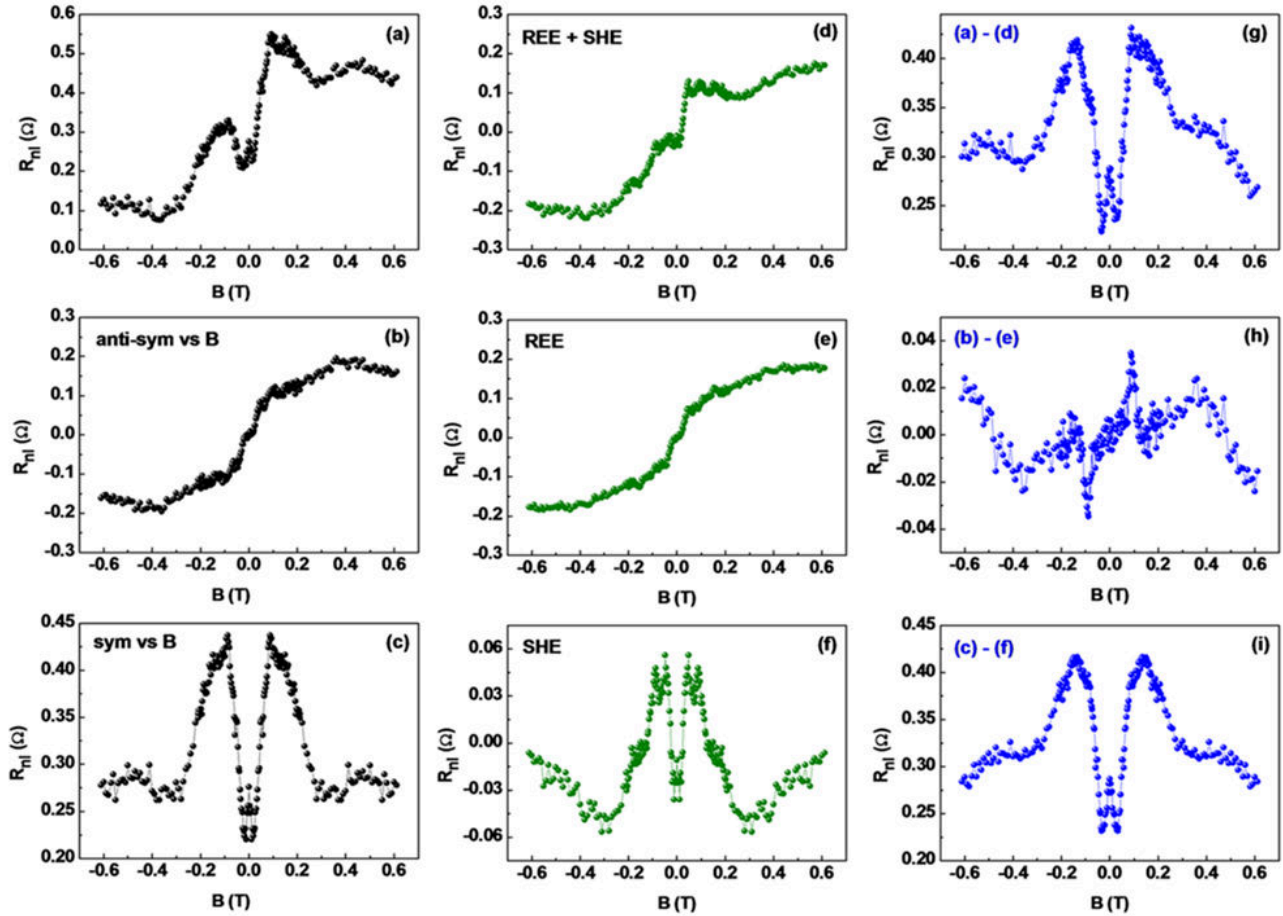

**Figure 4.** (a) Measured nonlocal resistance as a function of B for $\alpha = -15$ deg. Panels (b) and (c) show the anti-symmetric and symmetric vs B components, respectively, such that (a)=(b)+(c). Panels (e) and (h) are symmetric and anti-symmetric vs $\alpha$ components, respectively, for the component (b) such that (b)=(e)+(h). Similarly (c)=(f)+(i). Panel (d) shows the signal that includes both SHE and REE and was fitted by the model, (d)=(e)+(f). Panel (g) gives the total unknown component that does not obey the expected symmetries, (g)=(h)+(i).

## 4. Contact magnetization behaviour as a function of B

In Figure 5, we show the Hanle precession measurements performed with the in-plane magnetic field swept from $+0.6\,\mathrm{T}$ to $-0.6\,\mathrm{T}$ as the trace and from $-0.6\,\mathrm{T}$ to $+0.6\,\mathrm{T}$ as the retrace (B is applied under an angle of $\alpha \approx -15\,\mathrm{deg}$ in x-y plane). Before taking the trace a positive magnetic field of $+3\,\mathrm{T}$ is applied in order to align the contact magnetization. After crossing $B = 0\,\mathrm{T}$, the y-component of the magnetization of the contact switches to the opposite direction at $B \simeq 0.14\,\mathrm{T}$ which is seen as a jump in our signal indicated by the blue arrow. Similar behaviour is observed in the retrace with initial alignment of the contact magnetization at $-3\,\mathrm{T}$. The presence of these switches in the non-local resistance is another evidence of spin origin of our signals. Our detection circuit consists of two ferromagnetic electrodes which implies that in principle one should expect to see switches coming from both contacts whereas we observe only a single switch. Furthermore, our analysis (see Section 7) indicates that both contacts are spin sensitive and contribute to the spin transport detection with saturation fields along $x$ axis being $B1_{sat} \simeq 0.09\,\mathrm{T}$ and $B2_{sat} \simeq 0.3\,\mathrm{T}$ where the value for $B1_{sat}$ is extracted as a fitting parameter of the model. As suggested by these saturation fields Co contacts have different magnetic shape anisotropies which is supported by the difference in their geometrical aspect ratios (contact widths are $W1 = 0.7\,\mu\mathrm{m}$ and $W2 = 0.5\,\mu\mathrm{m}$). Therefore, we attribute the absence of the second switch in our measured signal to a non-ideal behaviour of the contact magnetization with small magnetic shape anisotropy.

For a clear interpretation for our analysis we select the positive field range of the trace curve and the negative field range of the retrace curve implying that for the full $B$ range in all the curves of both main manuscript and supplementary information (if not specified otherwise) the $y$ component of the contact magnetization is aligned with $y$ component of magnetic field.

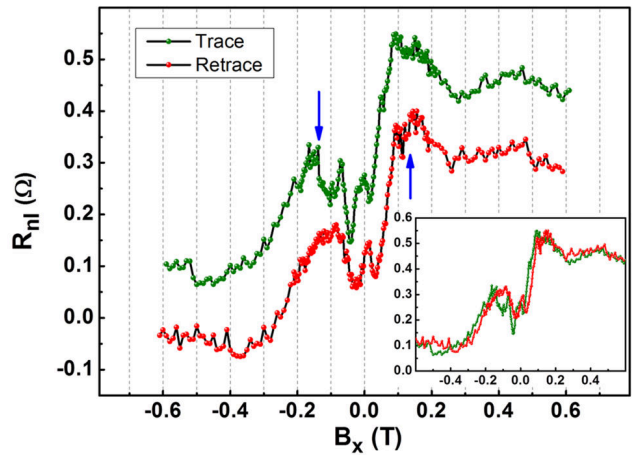

**Figure 5.** Trace and retrace of Hanle precession measurements with magnetic field applied in the x-y plane. The plots for the trace and retrace are shown with an offset in $y$ axis with respect to each other for clear demonstration of the switch of contact magnetization. The blue arrows determine the switching point. The inset shows the trace and retrace without the offset.

## 5. Model description

In order to describe our results qualitatively and understand the interplay between various contributions we use finite element analysis (COMSOL) for modeling the coupled spin-charge transport in the TMD-graphene device. Both Rashba-Edelstein and spin Hall effects are introduced phenomenologically and are assumed to have an origin from the proximity of TMD. In addition to that the model includes the tilt of the contact magnetization direction with the applied external magnetic field in accordance with the Stoner-Wohlfarth model. For in-plane orientation of magnetic field the contact magnetization is assumed to have a single easy axis within the sample plane. For the out-of-plane direction of magnetic field the contact magnetization is assumed to have an easy plane (parallel to the sample plane) with an easy axis within that plane.

To model diffusive spin transport in our device we numerically solve a set of spatially two-dimensional drift-diffusion equations for three polarization directions of non-equilibrium spin densities $n_s^x$, $n_s^y$, $n_s^z$ and electric potential $V$ ($n_s^i$ is expressed in electrical units of C/m$^2$).

The corresponding spin currents read as:

$$J_x^x = -D\frac{\partial n_s^x}{\partial x};$$

$$J_y^x = -D\frac{\partial n_s^x}{\partial y};$$

$$J_x^y = -D\frac{\partial n_s^y}{\partial x};$$

$$J_y^y = -D\frac{\partial n_s^y}{\partial y};$$

$$J_x^z = -D\frac{\partial n_s^z}{\partial x} + SH\frac{\partial V}{\partial y};$$

$$J_y^z = -D\frac{\partial n_s^z}{\partial y} - SH\frac{\partial V}{\partial x};$$

$$J_x^q = -\sigma_0\frac{\partial V}{\partial x} + ISH\frac{\partial n_s^z}{\partial y} + IREn_s^y;$$

$$J_y^q = -\sigma_0\frac{\partial V}{\partial y} - ISH\frac{\partial n_s^z}{\partial x} - IREn_s^x;$$

where, for example, $J_y^x$ is a current of spin aligned in $x$ direction flowing in the $y$ direction, $J_x^q$ and $J_y^q$ are $x$ and $y$ components of charge current. The diffusion coefficient $D$ and conductivity $\sigma$ of the system are connected through the Einstein relation $\sigma = De^2\nu$ with $\nu$ being density of states at the Fermi level. $SH$ and $RE$ are coefficients that determine the strength of induced spin Hall and Rashba-Edelstein effects. $ISH$ and $IRE$ are corresponding coefficients for inverse SHE and inverse REE and are chosen as $ISH = \frac{D}{\sigma}SH$ and $IRE = \frac{D}{\sigma}RE$ in order to obey reciprocity.

Using the continuity of the currents we formulate the diffusion equations for spin density and electric potential under an applied magnetic field $\vec{B} = (B_x; B_y; B_z)$:

$$-D\Delta n_s^x + \omega_z n_s^y - \omega_y n_s^z + \frac{n_s^x}{\tau_\parallel} = RE\frac{\partial V}{\partial y};$$

$$-D\Delta n_s^y + \omega_x n_s^z - \omega_z n_s^x + \frac{n_s^y}{\tau_\parallel} = -RE\frac{\partial V}{\partial x};$$

$$-D\Delta n_s^z + \omega_y n_s^x - \omega_x n_s^y + \frac{n_s^z}{\tau_\perp} = 0;$$

$$-\sigma_0 \Delta V = -IRE(\frac{\partial n_s^y}{\partial x} - \frac{\partial n_s^x}{\partial y});$$

where $\tau_\parallel$, $\tau_\perp$ are spin relaxation times for in and out-of-plane directions and spin precession is determined by $\omega_i = \frac{2\mu_B}{\hbar}B_i$.

It is important to note that SHE and REE are different in origin. REE creates a local spin density in the presence of the charge current while SHE creates a spin current which means that they enter the diffusion equations differently.

Gold contacts are incorporated into the model as regions with very high spin relaxation ($10000\tau_\parallel$) and large diffusion coefficient ($10000D$). To complete the problem the boundary conditions are defined as $V = 0$ V and $n_s^x = n_s^y = n_s^z = 0$ C/m² at the boundary of the contact used for the ground and $\vec{J^q} = \vec{J}_{input}$ and $\vec{J^x} = \vec{J^y} = \vec{J^z} = \vec{0}$ A/m, where $\vec{J}_{input}$ is an input normal to the contact boundary charge current supplied at the source contact. At all other boundaries of the system the current normal to the boundary (both spin and charge) is set to zero.

The geometry of the sample is copied from the optical images and implemented exactly into the geometry of the model, Figure 1(b) of the main text. From the measured carrier concentration dependence of the device conductivity $\sigma$ we extract the charge diffusion coefficient of the carriers $D$ (see Section 2), both of which are directly incorporated as fixed model parameters. Initial values and acceptable limits of spin relaxation times for in-plane and out-of-plane polarizations are taken from the available literature [2, 3] and later tuned in order to provide the best possible similarity to the complete set of the measured curves. Both ferromagnetic contacts are assumed to be spin sensitive and contribute to the measured spin signal.

The full set of equations is solved for each value of the applied magnetic field. Resulting maps of the spin density (Figure 2 of the manuscript) are then used to calculate an average spin density over the area underneath the Co contacts. The Stoner-Wohlfarth model is solved numerically for each field $B/B_{sat}$ applied under an angle $90\deg - \alpha$ to the magnetization easy axis. We calculate the projection of the spin density on the contact magnetization direction that is picked up in the measurement and obtain a non-local signal as a function of the applied field.

## 6. REE and SHE coefficients

One can rewrite the REE spin current density source as $RE\frac{\partial V}{\partial y} = n_s^x/\tau_\parallel$, where $n_s^x$ (or generalizing as $n_s$) is a uniformly generated spin density which is proportional to the charge current density $j_c = \sigma\frac{\partial V}{\partial y}$. It can be rewritten as $n_s = REj_c\rho\tau_\parallel$. In Ref. [4] the dimensionless REE efficiency $\alpha_{RE}$ is defined as the ratio between induced uniform spin density with respect to the supplied charge current density normalized by $2\upsilon_F$ ($\upsilon_F$ is Fermi velocity) which results in $\alpha_{RE} = 2\upsilon_F n_s/j_c = 2\upsilon_F RE\rho\tau_\parallel$. Under assumption for the value of contact polarizations $P1 = 0.3$ and $P2 = 0.2$ we extract the $RE$ coefficient from our experimental data to be $400\,\mathrm{S/m}$ at $E_F \simeq 130\,\mathrm{meV}$ which results in $\alpha_{RE} \simeq 2.8$. This value exceeds the theoretically predicted value that is limited by 1. However, we note here that as emphasized in Ref. [4] $\alpha_{RE}$ is not a suitable figure of merit for Fermi energies far from the Dirac point which is the case for our estimation given above. Moreover, our model assumes an idealistic behavior of the Co contact and sample which can result in an overestimated value for the efficiency. In addition to that, presence of many fitting parameters ($RE$, $\tau_\parallel$, $\tau_\perp$) that are coupled to each other results in an uncertainty that can again explain the discrepancy between our and theoretical estimations.

Note that the relation between defined here coefficient $SH$ and commonly used spin Hall angle $\theta_{SH}$ is $\theta_{SH} = SH/\sigma$. Here we extract $SH \simeq 0.13 \times 10^{-3}\,\mathrm{S}$ which results in $\theta_{SH} \simeq 0.13$ that is of similar order as reported in Ref. [5].

It is important to note here that in our particular case we are not able to extract the polarizations of the Co contacts independently from the strengths of REE and SHE effects. Therefore, the ratio of strengths of the two effect is the most reliable estimation that we can make as it does not depend on an average spin polarization of the contacts. For the extracted parameters we get $SH/RE = \frac{\theta_{SH}}{\alpha_{RE}}2\upsilon_F\tau \simeq 0.33\,\mu\mathrm{m}$.

## 7. Fitting of the measured results

Fitting of the measured results is done with the help of the optimisation module from COMSOL. The measured REE component is mostly determined by the tilt of the contact magnetization in plane meaning that the precessed spins (exploring the out-of-plane direction) give a minimal contribution to the measured signal due to relatively low spin relaxation time in-plane ($\tau_\parallel$ is in the order of few ps). In contrast the SHE contribution directly involves precession of spin from out-of-plane direction into in-plane direction and thus is sensitive to both spin relaxation times ($\tau_\parallel$ and $\tau_\perp$) and contact magnetization tilt. Therefore we use the SHE associated contribution of our data as an input for fitting procedure.

Furthermore, as discussed in Section 3 there is an unknown contribution present in our measurements that does not obey expected symmetries. Non-ideal behaviour of the contacts can not only significantly influence the value of the saturation field (perpendicular to the contact geometrical easy axis) but can also cause a tilt of an easy axis by an arbitrary angle. In this case one can expect an additional component coming from both SHE and REE that is neither symmetric nor anti-symmetric with respect to angle $\alpha = 0$ deg which results in an additional background after the data is anti-symmetrized with respect to $\alpha = 0$ deg. The larger the $\alpha$, the larger this background can be. Therefore, we conclude that the most reliable data for fitting is the SHE component extracted from the measurements with smallest available angle difference that is $\alpha = \pm 15$ deg (Figure 4h).

In Figure 6 we show the SHE contribution extracted from $\alpha = \pm 15$ deg measurements together with the produced fit. The extracted fitting parameters are $\tau_\parallel = 3.5$ ps, $\tau_\perp = 90$ ps, contact polarizations $P1 = 0.3$ and $P2 = 0.2$, in-plane saturation field of the softer contact $B1_{sat} = 0.09$ T (the saturation field of the harder contact $B2_{sat}$ is assumed to be 0.3 T which is estimated from the saturation level of the measurement taken at $\alpha = 0$ deg) and the background level $A = -0.02\,\Omega$. The margins for the given parameters are relatively large and therefore the extracted values should only be seen as estimations. In order to illustrate the influence of each parameter on the final fit we deviate its value within a certain margin and calculate the deviating dependences which are shown in Figure 7.

The extracted SHE component has a relatively small background that is likely to come as a result of (anti-)symmetrization procedures in the presence of an unknown component. In order to provide a better fit from the model we introduce a constant background $A = -0.02\,\Omega$.

It is apparent from Figure 7 that several parameters are heavily coupled with each other. To narrow down the range of possible parameters for the fitting procedure we used following restrictions that are based on the experience of our group and on the results published by other groups: $P1 \leq 0.4$, $P2 \leq 0.4$, $0.5\,\text{ps} \leq \tau_\parallel \leq 5\,\text{ps}$ and $5 \leq \tau_\perp/\tau_\parallel \leq 40$.

The size of the measured signal determines only the product of $RE * P$ ($P$ is average polarization of Co contacts) and therefore we cannot independently determine them. For the final estimation of $RE$ and $SH$ coefficients we assume that $P1 \simeq 0.3$ and $P2 \simeq 0.2$. Furthermore, the ratio $P1/P2$ and parameters $\tau_\perp$ and $\tau_\parallel$ are also strongly coupled which means that one can obtain a satisfactory fit to the same data for the full range of $2\,\text{ps} \leq \tau_\parallel \leq 5\,\text{ps}$ (where 5 ps is the limit of the accepted range for the in-plane spin relaxation time) by adjusting the other

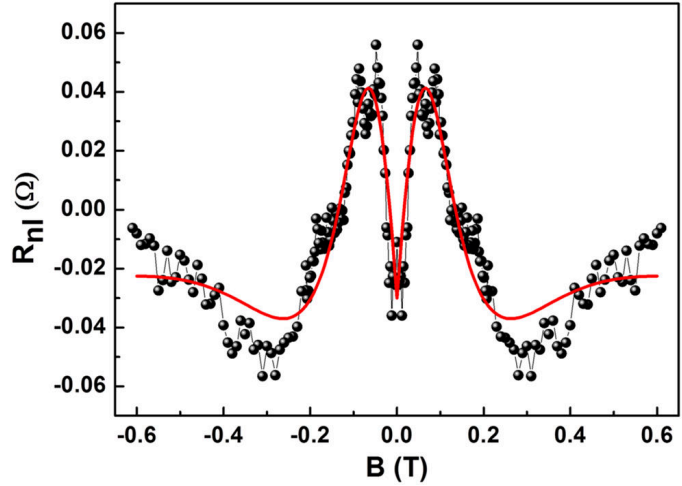

**Figure 6.** SHE component of the signal measured at $\alpha = -15\,$deg together with a fit from the model.

parameters accordingly.

Note that the in-plane saturation field for one of the contacts $B1_{sat}$ was found to be around $0.09\,$T which is not very typical, as the values obtained from different experiments [2] for similarly shaped contacts are in the range between $0.13\,$T and $0.3\,$T. However, it is apparent from Figure 7 that the position of the peak at $B \simeq 0.07\,$T is mostly sensitive to the value of $B1_{sat}$ (Figure 7a). This means that $B1_{sat}$ is very well defined by the fitting procedure and the value of, for example, $0.13\,$T would not give a satisfactory agreement with the measured data. This is in agreement with the presence of an unknown component associated with non-expected behavior of the contact magnetization.

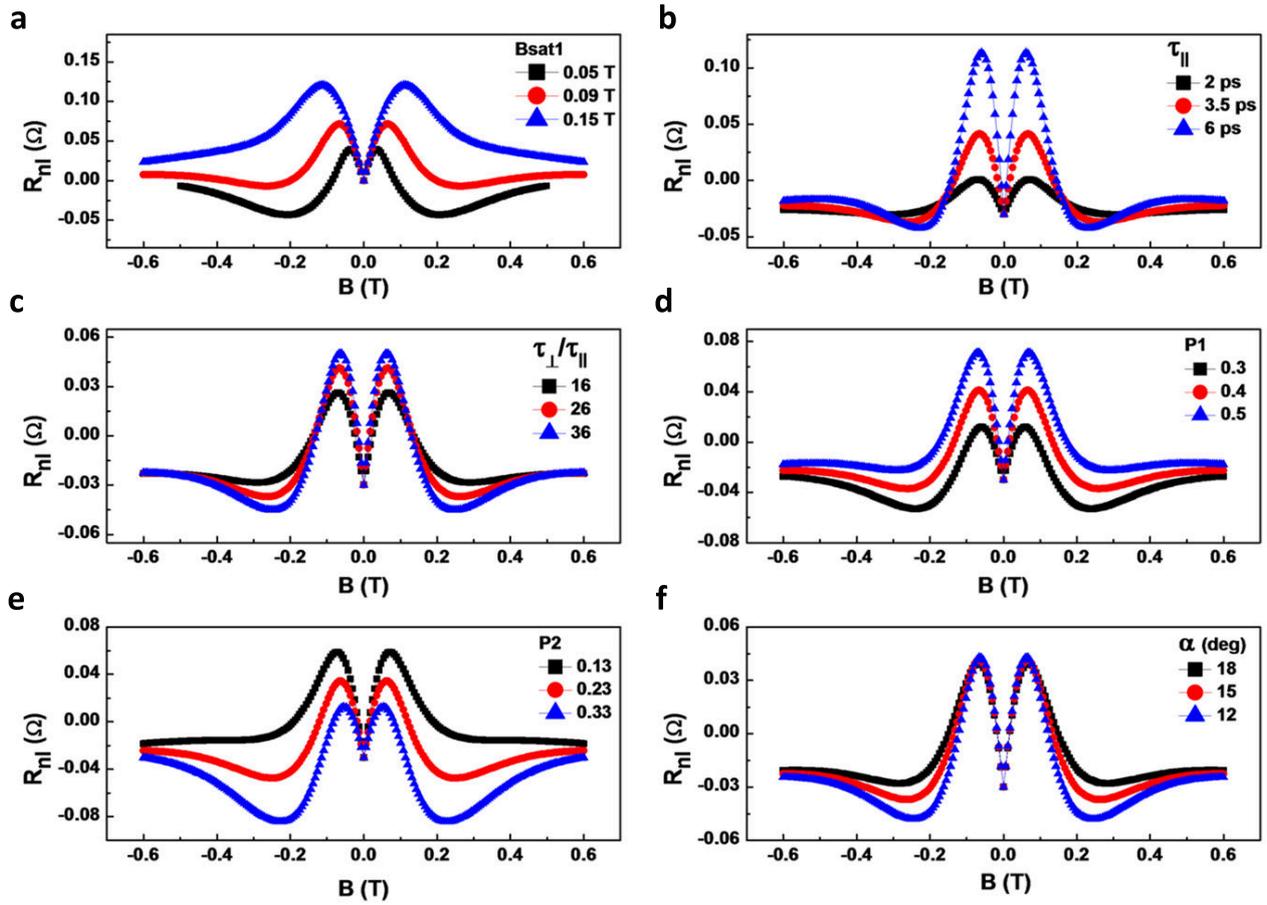

**Figure 7.** Panels (a)-(f) give the deviations of the fitting curve for each of the used fitting parameters. The red curve is the same in all the panels and gives the resulting fit shown in Figure 6.

## 8. y-z plane Hanle precession measurements

By applying an out-of-plane magnetic field we can acquire additional information on the system parameters. In order to do that we reloaded the sample into a different setup where the out-of-plane rotation is accessible along the $y - z$ plane. We have measured the non-local signal as a function of magnetic field B applied along the direction oriented with angle $\varphi$ with respect to the $z$ axis. In Figure 8 we show the measured dependences at $\varphi = 0, \pm 45$ deg. Following the same logic as in section 9 we perform the anti-symmetrization procedure with respect to both B and $\phi$ and obtain a signal that here is much smaller in magnitude than any other signal measured in any other geometry. We associate this difference with the fact that the sample parameters changed significantly during the reloading procedure (mostly due to doping by water) which results in an undistinguishable from noise spin related signal.

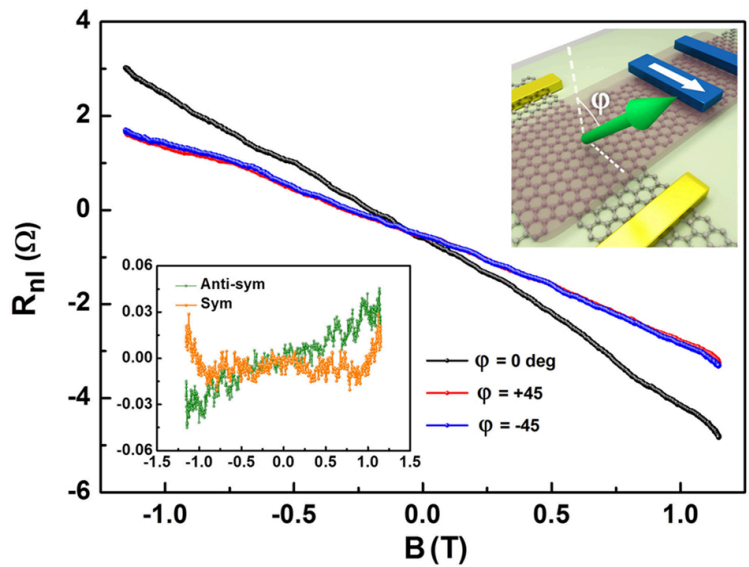

**Figure 8.** $y - z$ plane Hanle precession measurements. The inset is the symmetric and anti-symmetric components of the subtraction of the measurements performed at $\varphi = \pm 45$ degrees.

## 9. x-z plane Hanle precession measurements and modeling

In order to confirm the presence of both in-plane and out-of-plane spin polarization directions we reloaded for the second time our sample into another holder. In this holder magnetic field can be applied in any direction along the $x-z$ plane with $\theta$ being an angle between the magnetic field direction and $z$ axis. The measurements were performed both at room temperature and $10\,\mathrm{K}$. Room temperature measurements are shown in Figure 5 of the manuscript and $10\,\mathrm{K}$ measurements are presented here in Figure 9a. Having a $B_z$ component of magnetic field one should expect additional components coming from Hall effect and magnetoresistance (MR). The non-local geometry of our measurements insures that charge current induced contributions are minimal which is confirmed by the low background level in all our measurements. However, due to the Ohmic spreading of the charge current there is a finite local current flowing between Co contacts along $y$ axis which produces a measurable small Hall effect component. In order to exclude MR and Hall effect completely we again employ appropriate (anti)symmetrization procedures. Firstly, to eliminate MR effect (or any other symmetric in $B_z$ contributions) from our analysis we anti-symmetrize our measured dependences with respect to the applied B (main panel of Figure 9a). Further, SHE and Hall effect should give the same contributions for opposite angles $+\theta$ and $-\theta$ since the $B_z$ component is the same. Therefore, anti-symmetrization of the data with respect to the angle $\theta$ should result in a component (shown in the inset) that can only be explained by REE. In the panel (b) we show the corresponding calculated contributions coming from REE for both angles $30\,\mathrm{deg}$ and $60\,\mathrm{deg}$. The curves are calculated using the spin transport parameters extracted from the in-plane measurements. These curves qualitatively resemble similar behaviour as experimental data, however, the magnitude is different most likely due to the change in the sample parameters during the reloading procedure.

It is known that the in-plane saturation field of the contact magnetization direction is much smaller than that for the out-of-plane. Therefore, one should expect that with magnetic field applied at an angle $\theta$ in the $x-z$ plane the contact magnetization should first mostly rotate within the $x-y$ plane until its tilting in this plane is saturated along $x$. After that the magnetization starts to tilt from the $x-y$ plane towards the $z$ axis within the $x-z$ plane. Under this assumption the saturation point in the measured curves should be determined by $B_x$ (saturation happens at $B_x = B_{sat}$, where $B_{sat}$ is a saturation field within the $x-y$ plane, along $x$ direction). Our experimental observation of saturating behaviour as seen from Figure 9a (inset) is consistent with the described logic and is reproduced in the modeled dependences. Nevertheless, the behaviour beyond the saturation point is almost flat in the experimental data but is changing considerably in the calculated dependences. In order to explain these differences one should consider that the sample went through the procedure of unloading/loading from/into the setup which is known to significantly influence both the properties of the channel and of the contacts whereas we used the same parameters for modeling the out-of-plane dependences as extracted from the in-plane measurements. Indeed, remeasured dependence of the channel resistance versus gate voltage (see Sec. 3) suggests a shift of the Fermi level in the sample from electron to hole-doped regime indicating major changes in the system properties presumably implying changes in all the relevant parameters including $RE$ (which is observed to depend on the carrier density).

With an understanding of the origin of the saturation points in both measured and calculated data we can conclude that spin signal before the kink is mostly determined by the in-plane rotation of the contact magnetization and beyond the kink it is mostly determined by the precession of the spins out-of-plane where it is picked up by the contacts with tilted out-of-plane magnetization. Therefore, by tuning appropriately the anisotropy level we can significantly flatten out the calculated dependences after the kink as it is shown in the inset of Figure 9b. This again points towards the change in the sample parameters after reloading it into the setup.

In conclusion, with out-of-plane applied field we also observe a component that can only be associated with REE, however, the exact behaviour and the magnitude of the measured signal differs from the calculated dependences most likely due to the change in the sample parameters during the sample reloading procedure.

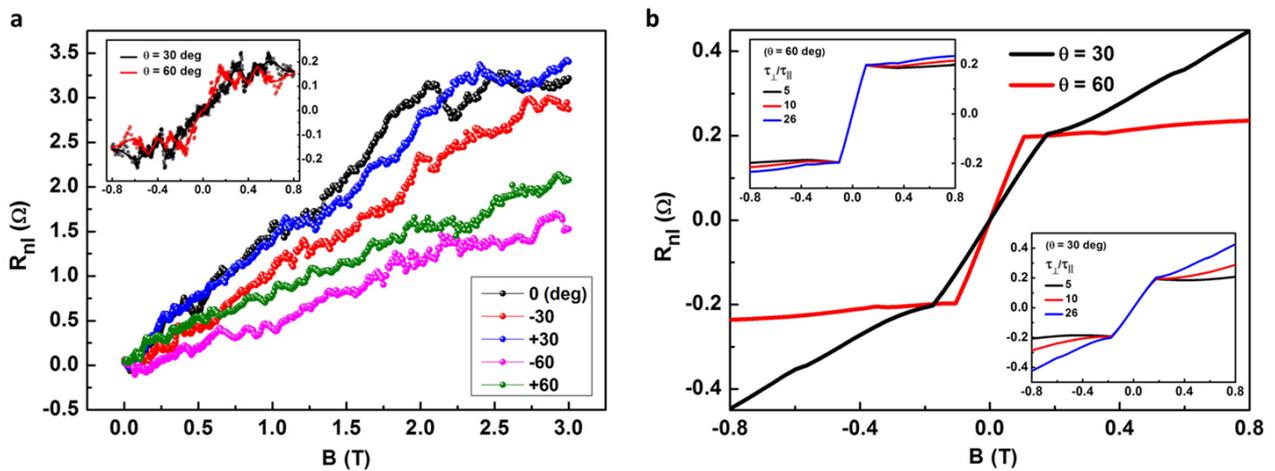

**Figure 9.** (a) Anti-symmetrized with respect to B non-local resistance measured for different directions of the applied field. Inset: measurements as in main panel but further anti-symmetrized with respect to $\theta$. (b) Calculated non-local signal with spin transport parameters extracted from the in-plane measurements. Inset: same as in the main panel with having the anisotropy ratio $\tau_\perp/\tau_\parallel$ varied between 5, 10 and 26 for $\theta = 30$ and 60 degrees.

## 10. Second harmonic measurements

Our measurements are performed by using a lock-in amplifier technique which allows to separate linear and non-linear response of the system from each other. All our results presented elsewhere in this manuscript and supporting information are the linear response of the sample detected as the first harmonic signal. However, along with the first harmonic signal we also detected the second harmonic signal which is seen to give a clear Hanle-like dependence over the measured $B$ range. In Figure. 10 we show the measured second harmonic signal at different gate voltages with magnetic field applied along $x$ axis.

There are two possible reasons for the non-trivial second harmonic signal to appear in our system. First of all, a thermal gradient that originates from the Joule heating can result in a finite spin density due to spin Nernst and/or spin Seebeck effects. This means that a thermal gradient in the $x$ direction can produce spin current in $y$ direction with spin polarization in $z$ direction (analogous to SHE in linear response) and spin density polarised in $y$ direction (analogous to REE in linear response). Spin current in $y$ direction results in a spin density of opposite polarization at opposite edges of the sample and cannot be detected in our measurement geometry since the spin sensitive contact averages the spin potential over the full width of the device. On the other hand, uniform spin density polarised along $y$ axis can be detected and should result in a symmetric vs $B$ Hanle curve. However, in this case we should be able to distinguish clearly the parallel (P) and anti-parallel (AP) (with respect to the uniform induced spin density) states of the contact magnetization resulting in two P and AP signals of the same magnitude but with opposite sign in a similar fashion as regular P and AP Hanle curves in only graphene samples. In our measurements we are not able to clearly identify the P and AP states and thus we cannot explain our results with the described above model.

An alternative explanation of the measured dependences is a non-linear detection of the spin density which is based on spin dependent conductivity in the presence of non-zero spin accumulation [6]. Following this mechanism the spin densities created in linear response of the system can still be detected in the second harmonic with the difference that the detection is not sensitive to the sign of the spin density but only to its magnitude squared. This logic also results in a symmetric in $B$ Hanle curves, however, at this stage we are not able to correlate the non-linear spin density detection with the measured dependences.

In conclusion, our measurements clearly suggest a presence of non-linear response of our sample which can either be associated with thermal gradient driven effects and/or non-linear spin detection mechanism. However, we are unable to clearly identify which of the two is responsible for the measured results strongly suggesting further measurements and analysis which will be presented in future publication.

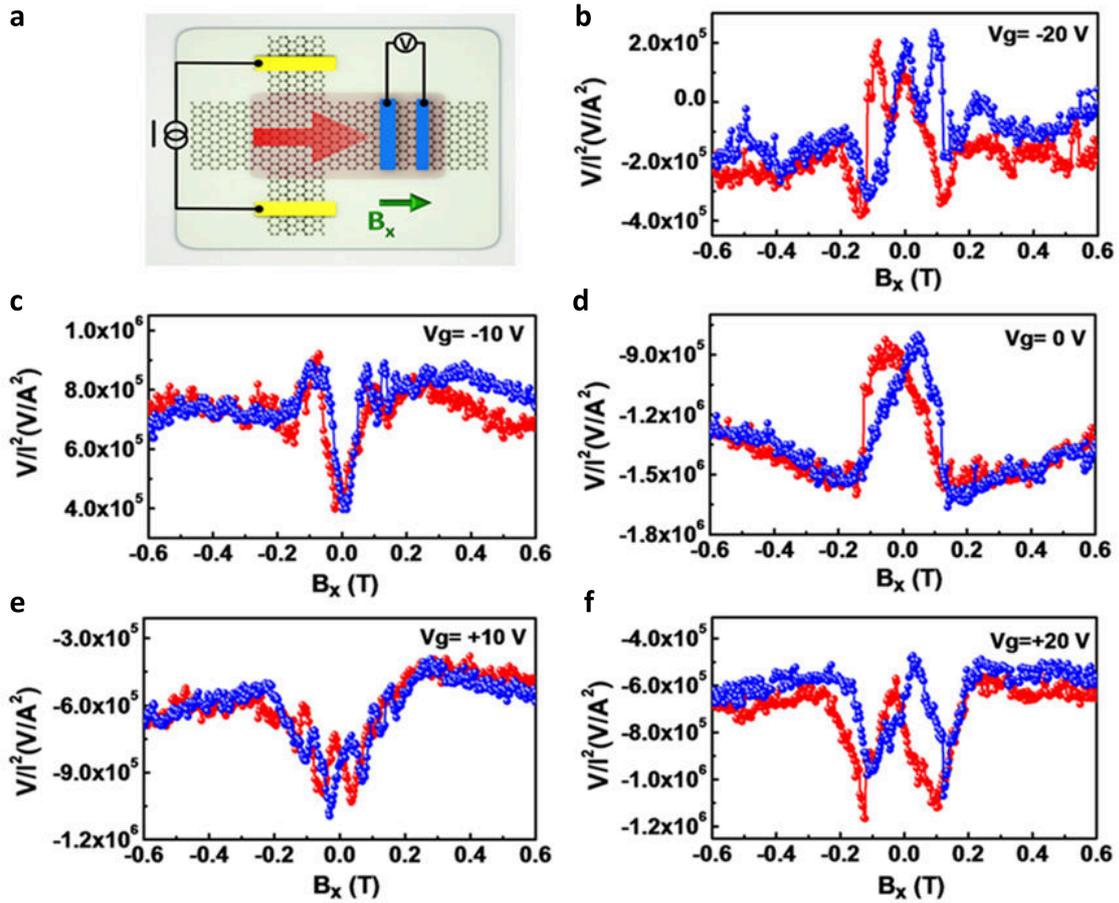

**Figure 10.** (a) Sketch of the device and the measurement geometry for the second harmonic measurements. The red arrow represents the thermal gradient $(-\nabla T)$. (b)-(f) Modulation of the second harmonic signal $(V/I^2)$ as a function of $B_x$, measured at different gate-voltages.